# Investigation of Novel Preclinical Total Body PET Designed With J-PET Technology: A Simulation Study

M. Dadgar , S. Parzych, F. Tayefi Ardebili, J. Baran, N. Chug, C. Curceanu, E. Czerwiński, K. Dulski , K. Eliyan, A. Gajos , B. C. Hiesmayr , K. Kacprzak , Ł. Kapłon , K. Klimaszewski , P. Konieczka, G. Korcyl , T. Kozik , W. Krzemien, D. Kumar, S. Niedzwiecki , D. Panek, E. Perez del Rio , L. Raczyński , S. Sharma , Shivani , R.Y. Shopa , M. Skurzok, K. Tayefi Ardebili, S. Vandenberghe , W. Wislicki, E. Ł. Stępień´ , and P. Moskal

*Abstract*—The growing interest in human-grade total body positron emission tomography (PET) systems has also application in small animal research. Due to the existing limitations in human-based studies involving drug development and novel treatment monitoring, animal-based research became a necessary step for testing and protocol preparation. In this simulation-based study two unconventional, cost-effective small animal total body PET scanners (for mouse and rat studies) have been investigated in order to inspect their feasibility for preclinical research. They were designed with the novel technology explored by the Jagiellonian-PET (J-PET) Collaboration. Two main PET characteristics: sensitivity and spatial resolution were mainly inspected to evaluate their performance. Moreover, the impact of the scintillator dimension and time-of-flight on the latter parameter was examined in order to design the most efficient tomographs. The presented results show that for mouse TB J-PET the achievable system sensitivity is equal to 2.35% and volumetric spatial resolution to $9.46 \pm 0.54$ mm$^3$, while for rat TB J-PET they are equal to 2.6% and $14.11 \pm 0.80$ mm$^3$, respectively. Furthermore, it was shown that the designed tomographs are almost parallax-free systems, hence, they resolve the problem of the acceptance criterion tradeoff between enhancing spatial resolution and reducing sensitivity.

## I. INTRODUCTION

Biomedical research on animals is an inherent preceding part of human-based studies. This fact is dictated by the necessary safety restrictions in for example testing phase of drug development or novel treatment monitoring [1]. Since rodents like mice and rats share with humans a large number of common diseases, these small animals became a suitable alternative system for preclinical research [2]. Moreover, the high reproduction rates of rodents provide accessible experimental and control populations for statistical studies [3].

Positron emission tomography (PET) is a typical choice for medical research on small animals. It enables noninvasive imaging of the temporal and spatial distribution of the radiopharmaceutical in vivo, which found use in the early oncological diagnosis, physiological imaging, as well as treatment monitoring [3]. Due to the fact that rodent-based studies involve animals of sizes several times smaller than human scale, the human-grade PET systems are not adequate for them. The small sizes of anatomical structures force the utilization of dedicated small animal PET tomographs characterized by the higher spatial resolution and much more compact dimensions.

Manuscript received 1 July 2022; revised 26 August 2022; accepted 19 September 2022. Date of publication 4 October 2022; date of current version 3 February 2023. This work was supported in part by the Foundation for Polish Science through TEAM POIR.04.04.00-00-4204/17; in part by the National Science Centre (NCN), Poland, under Grant 2021/42/A/ST2/00423; in part by PRELUDIUM 19 under Agreement UMO-2020/37/N/NZ7/04106; in part by the Ministry of Education and Science under Grant SPUB/SP/530054/2022; in part by the grant from the SciMat and qLife Priority Research Areas through the Strategic Programme Excellence Initiative at the Jagiellonian University; in part by the Jagiellonian University under Grant CRP/0641.221.2020; and in part by EU Horizon 2020 Research and Innovation Programme, STRONG- 2020 Project under Grant 824093. *(Corresponding author: M. Dadgar.)* This work did not involve human subjects or animals in its research. M. Dadgar, S. Parzych, F. Tayefi Ardebili, J. Baran, N. Chug, E. Czerwiński, K. Dulski, K. Eliyan, A. Gajos, K. Kacprzak, Ł. Kapłon, G. Korcyl, T. Kozik, D. Kumar, S. Niedźwiecki, D. Panek, E. Perez del Rio, S. Sharma, Shivani, M. Skurzok, K. Tayefi Ardebili, E. Ł. Stępień´, and P. Moskal are with the Faculty of Physics, Astronomy, and Applied Computer Science and Total Body Jagiellonian-PET Laboratory, Jagiellonian University, 31-007 Kraków, Poland (e-mail: meysam.dadgar@uj.edu.pl; p.moskal@uj.edu.pl). C. Curceanu is with the Faculty of Physics, Astronomy, and Applied Computer Science and Total Body Jagiellonian-PET Laboratory, Jagiellonian University, 31-007 Kraków, Poland, and also with INFN, Laboratori Nazionali di Frascati, 00044 Frascati, Italy. B. C. Hiesmayr is with the Faculty of Physics, University of Vienna, 1010 Vienna, Austria. K. Klimaszewski, P. Konieczka, L. Raczyński, R.Y. Shopa, and W. Wiślicki are with the Department of Complex Systems, National Centre for Nuclear Research, 05-400 Otwock-Świerk, Poland. W. Krzemień´ is with High Energy Physics Division, National Centre for Nuclear Research, 05-400 Otwock-Świerk, Poland. S. Vandenberghe is with the Department of Electronics and Information Systems, MEDISIP, Ghent University-IBiTech, 9000 Ghent, Belgium. Color versions of one or more figures in this article are available at https://doi.org/10.1109/TRPMS.2022.3211780. Digital Object Identifier 10.1109/TRPMS.2022.3211780

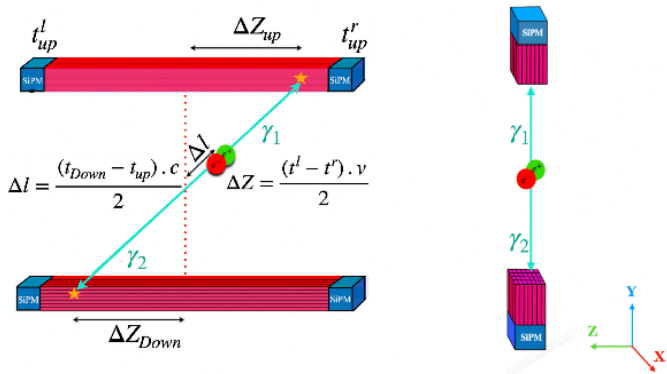

Fig. 1. (Left) Gamma quanta from electron–positron annihilation interact with plastic scintillators (red). These interactions produce light signals which can be detected by SiPM (blue) located at each end of the scintillator. By knowing the speed of light signals (v) and arrival time difference to each SiPM, the interaction position along plastic strips can be calculated [19]. (Right) Schematic view from detection of gamma quanta caused by annihilation in traditional PET scans.

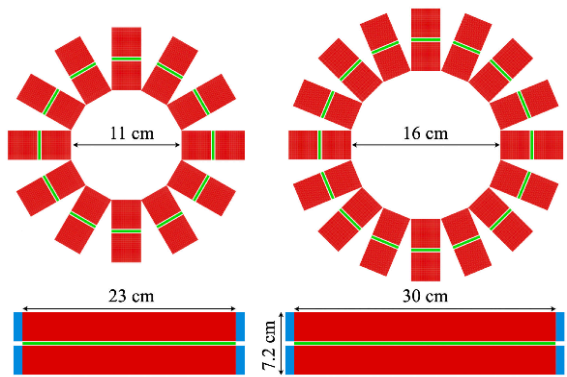

Fig. 2. Graphical visualization of the mouse TB J-PET (left) and rat TB J-PET, detection units called panels (red) in the form of the two detection layers with 5-mm gaps between them, coupled with SiPM at each end (blue). Each module is constructed of 32 × 32 × 1 array of EJ-230 plastic scintillators. To enhance axial resolution, among every two modules and right between layers the array of WLS strips perpendicular to the plastic scintillator has been located (green).

Despite the exceptional spatial resolution, small animal scanners suffer from poor sensitivity—another key characteristic in the PET performance. This directly transfers to the lower image quality and necessity of longer scans with higher administered radioactive doses. Since there are regulations and limits on the latter, there is a demand for the development of new tomographs able to perform high-resolution imaging with substantial sensitivity. Recent years, in the field of PET, are characterized by the investigations and developments of new category of human- grade scanners, namely, the total body systems [4], [5], [6], [7], [8]. Their greater coverage over the human body with respect to the standard clinically available tomographs not only enhances the scanner's sensitivity but also enables the dynamic imaging [9]. The advantages of the total body systems for humans brought attention and interest in extending of the field- of-view in small animal PETs as well [10].

Nevertheless, the development of a total body PET comes with new challenges, mainly due to the multiplied number of utilized detectors and electronics [9]. One of the solu- tions for such design was presented by the Jagiellonian-PET (J-PET) Collaboration from Jagiellonian University in Cracow, Poland, [11], [12], [13], [14], [15], [16]. Their novel approach to the tomograph construction based on the organic scintilla- tors was proven to be a viable, cost-effective alternative [17]. In oppose to the traditional use of crystal scintillators, the J-PET design utilizes axially oriented plastic scintillator strips arranged in the portable detection units with silicon photomultiplier readout situated on both ends (as shown in Fig. 1) [13], [18].

This simulation-based study is dedicated to checking the applicability of the organic plastic scintillators in the PET studies on small animals, especially in the context of the optimization of spatial resolution by varying scintillator cross section. Furthermore, the feasibility of the resulting small animal total body PET tomographs designed with the J-PET technology is investigated. For the evaluation of their performance, two of the main characteristics of scanners: sensitivity and spatial resolution have been investigated. Additionally, the influence of the angular acceptance criterion, which was shown to enhance the performance of human-grade total body systems has been examined [20].

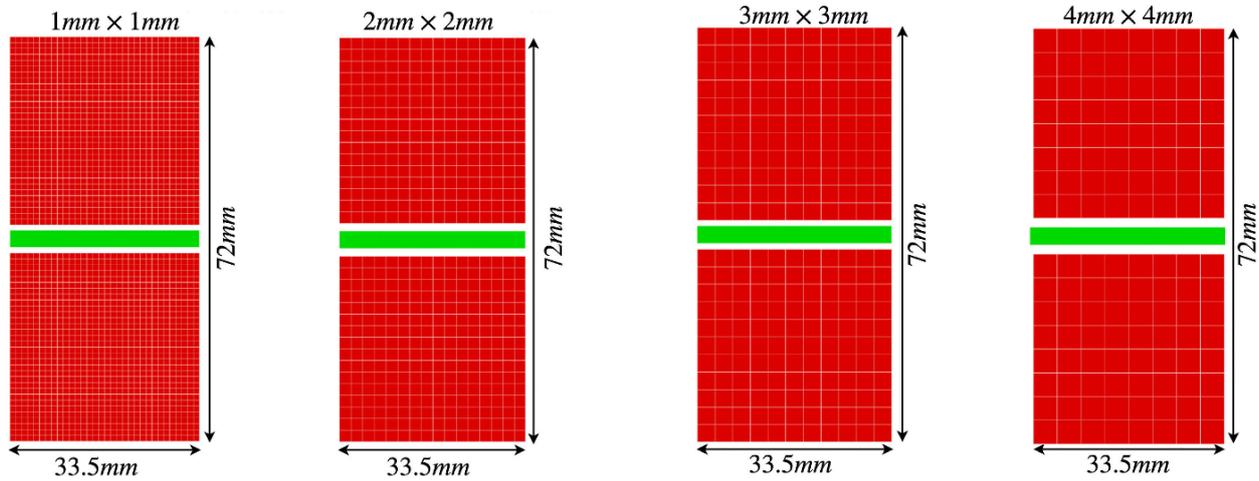

Fig. 3. Four panel types constructed from scintillators of 1 mm × 1 mm (upper-left), 2 mm × 2 mm (upper-right), 3 mm × 3 mm (lower-left), and 4 mm × 4 mm (lower-right) cross sections.

## II. METHODS

In this study, a small animal total body system adjusted for the mice's anatomical size and designed with the J-PET technology was investigated (see Figs. 2 and 3 top-left). This 23-cm long tomograph with 11 cm of diameter will be from now on referred to as mouse TB J-PET. It is composed of axially arranged EJ-230 "Eljen Technology" plastic scintillation strips coupled on both ends with silicon photomultipliers (SiPMs) [21] and grouped into two layers of 33.5 mm × 33.5 mm × 230 mm modules. Two consecutive modules in a radial direction constitute a panel. In total, it consists of 12 panels. Moreover, in order to improve the axial resolution, a layer of wavelength shifting (WLS) strips arranged perpendicularly to the scintillator strips is incorporated into this TB J-PET [18], [22].

The presented research has been carried out with the use of the Geant4 Application for Tomographic Emission (GATE) software [23], [24], [25]. GATE is a validated open-source simulation toolkit dedicated to numerical simulations in nuclear medicine and imaging systems. The GATE software was used to simulate the aforementioned total body small animal J-PET systems proposed by the J-PET Collaboration. The obtained hit-based results were further analyzed by the GOJA software, specifically, designed for J-PET-like geometries [26]. Due to the fact, that in the case of organic scintillators, the Compton scattering is the dominating channel for annihilation photons interaction, the traditional energy window is exchanged for the energy threshold of 200 keV [27]. Since the spatial resolution is closely related to the scintillator dimensions a study of the strips cross section has been performed taking into account parameters, such as sensitivity, spatial resolution, and time-of-flight (TOF) resolution. For that four cross sections were considered of one scintillator in a 33.5 mm × 33.5 mm quadratic array (as shown in Fig. 3): 1 mm × 1 mm, 2 mm × 2 mm, 3 mm × 3 mm, and 4 mm × 4 mm were evaluated.

The sensitivity of scanners represents their ability for photon detection. It consists of many factors like geometrical coverage of detectors, their efficiency at 511 keV, etc., [28]. The system sensitivity is defined as a fraction of detected events to all produced ones [29]. In order to estimate system sensitivities of four inspected mouse TB J-PET tomographs, a 370-kBq back-to-back point source without any surrounding phantom has been simulated in the axial and radial center. Moreover, for the sake of this and all subsequent simulations the energy threshold of 200 keV [18], [27] and coincidence window of 3 ns have been used.

The reconstruction of simulated data was performed with the OSEM iterative algorithm provided by the quantitative emission tomography iterative reconstruction (QETIR) software [30], [31]. The images were reconstructed using ten iterations with two subsets each while taking into account all types of the coincidences. The sensitivity map has been generated by the QETIR software with 1mm×1mm×1mm voxel size.

Nowadays, the TOF technique is utilized to enhance the performance of tomographs. It allows for localizing annihilation points with more precision than in the case of non-TOF PET scanners and therefore improves the spatial resolution. To investigate the effect of the various TOF resolutions for different dimensions of plastic scintillator, a 1-MBq back-to-back point source has been simulated at the center of the scanner.

The spatial resolution as one of the most important characteristics of PET scanners reflects the dimensions of distinguishable lesions, which is of special importance in the case of small animal studies [32]. Its quality was inspected as the full width at half maximum (FWHM) of the reconstructed centrally located 1-MBq back-to-back point source without any surrounding phantom.

Furthermore, after estimating the best performing configuration of the mouse TB J-PET, a more detailed study of the chosen system has been performed. Additionally, a second small animal total body scanner dedicated to rat studies was designed in order to adjust to the anatomical size difference between mice and rats (see Fig. 2) [33]. It consists of 16 panels as defined for the previous tomograph (however with a 30-cm axial length) which provide 30 cm of axial-field- of-view (AFOV) and 16 cm of diameter. From now, on this system will be referred to as rat TB J-PET. A more detailed geometrical configuration can be found in Table I.

TABLE I
GEOMETRICAL CONFIGURATIONS OF SMALL ANIMAL TB J-PET TOMOGRAPHS

| Scanner | Mouse TB J-PET | Rat TB J-PET |
|---|---|---|
| Scintillator materials | EJ-230 | |
| Scintillators dimensions [mm] | 1x1x230 | 1x1x300 |
| Transverse scintillator pitch [mm] | 0.05 | |
| Number of panels | 12 | 16 |
| Number of the modules | 24 | 32 |
| Number of detection layers | 2 | |
| Gaps between layers [mm] | 5 | |
| Diameter [cm] | 11 | 16 |
| AFOV [cm] | 23 | 30 |

TABLE II
GEOMETRY-BASED CHARACTERISTICS OF SMALL ANIMAL TB J-PET TOMOGRAPHS

| Scanner | Mouse TB J-PET | Rat TB J-PET |
|---|---|---|
| Solid angle coverage [rad] | 11.34 | 11.09 |
| Packing fraction [$\varphi$] | 0.96 | 0.96 |
| Geometric efficiency [rad] | 10.84 | 10.6 |

The geometric efficiency of the proposed tomographs were estimated according to the formula [34]

$$\Xi = \Omega \varphi \quad (1)$$

where ($\Omega$) is the overall solid angle coverage and ($\phi$) is the packing fraction. They are calculated as

$$\Omega = 4\pi \sin\left[\tan^{-1}(A/D)\right] \quad (2)$$

where $A$ is the axial length of the tomograph, $D$ its diameter and

$$\varphi = \frac{wh}{(w + d_w)(h + d_h)}. \quad (3)$$

Packing fraction is a dimensionless parameter as shown in (3), where $w$ is the width of the detector module, $h$ is its axial height, and $d_w, h$ is the corresponding dead space. The obtained values are presented in Table II.

In order to evaluate the performance of the final two scanners more thorough studies of sensitivity, spatial resolution, and event selection have been performed [35], [36].

So as to estimate sensitivities possible to obtain within the whole inspected small animal TB tomographs, back-to-back point sources of 370-kBq activity each and without any surrounding phantoms have been simulated with different axial and radial offsets (see Table III).

TABLE III
POSITIONS OF POINT SOURCES FOR SENSITIVITY ESTIMATION

| Scanner | Mouse TB J-PET | Rat TB J-PET |
|---|---|---|
| Radial offset [mm] | 0, 5, 10, 15, 25 & 40 | 0, 5, 10, 15, 25, 50 & 60 |
| Axial offset [mm] | 0 to 80 By the steps of 10 | 0 to 120 By the steps of 10 |

The spatial resolution was inspected along each of the radial, tangential, and axial directions [37]. For each TB tomograph, four back-to-back point sources without any surrounding phantoms and situated in the (1/2) and ~ (1/4) of axial and radial directions have been simulated.

It was previously shown [20] that human-grade total body PET scanners suffer from detection of oblique lines of responses (LORs), which have a negative influence on the quality of the reconstructed image. Notably, the spatial resolution is degrading due to the strong photon attenuation in the body and parallax error [38]. A solution for that problem was presented by introducing an acceptance criterion [20], [39], [40], [41] during the event selection, which sets angular limits on the LORs taken into image reconstruction. For the human total body PET tomography, it was presented that the $45^\circ$ acceptance angle cut can be the optimal selection for resolution improvements [20]. This study will investigate if the acceptance parameter is as before a valid criterion for small animal total body scanners.

## III. RESULTS

In order to design a novel small animal total body PET tomograph, which among other characteristics requires high spatial resolution, a detailed investigation of the effect of the plastic scintillator's transverse dimensions on the scanner's performance has been performed. For this test, the mouse TB J-PET has been chosen as a PET system. While keeping the geometrical configuration as previously described, only the cross section of the plastic strip was varied (see Fig. 3).

First, the influence on sensitivity was estimated based on a simulation of a centrally located point source. The obtained values were equal to 2.38%, 2.40%, 2.66%, and 2.45% for 1 mm, 2 mm, 3 mm, and 4 mm versions, respectively. The slow growth in sensitivity between 1, 2, and 4 mm is caused by the reduction of the number of in between scintillator gaps. The deviating result for the 3 mm × 3 mm cross section comes from the additional 2-mm thickness of material in contrast to the rest of inspected types. Such addition was necessary due to the not natural division of the module into 3-mm sections. The results indicate that while keeping the same amount of scintillator material, there is almost no difference in sensitivity for various cross sections.

To assess the impact of TOF resolution on the spatial resolutions of each scintillator cross section, a centrally located point source was simulated (see Fig. 4). In the case of radial resolution for TOF above 100 ps, there is no significant difference, especially for smaller strip dimensions. However, this tendency is not repeated in axial resolution where the dependence is more linear. In the following studies, 230-ps FWHM has been chosen as TOF resolution, consistent with previous work [42]. To evaluate the spatial resolution, a point source at the center of the mouse TB J-PET scanner was simulated. Due to the central location of the source and symmetrical configuration of the tomograph, only radial and axial resolutions were investigated (see Fig. 5). There is a clear dependence of the transverse dimension of scintillator strips on the obtainable FWHM values. The best achievable resolution in both axial and radial directions is provided by the smallest cross section. Moreover, a further decrease in the spatial resolution can be obtained with a number of reconstruction iterations. In the radial direction, the plateau is reached around 6th iteration, while in the axial direction around 8th.

The aforementioned results show that the 1 mm × 1 mm cross section of the plastic scintillator strip ensures the best performance of the tomograph. Based on it, the corresponding rat and mouse total body J-PET systems have been taken into consideration for a thorough study of their performance. The geometrical configuration of the two scanners can be found in Table I.

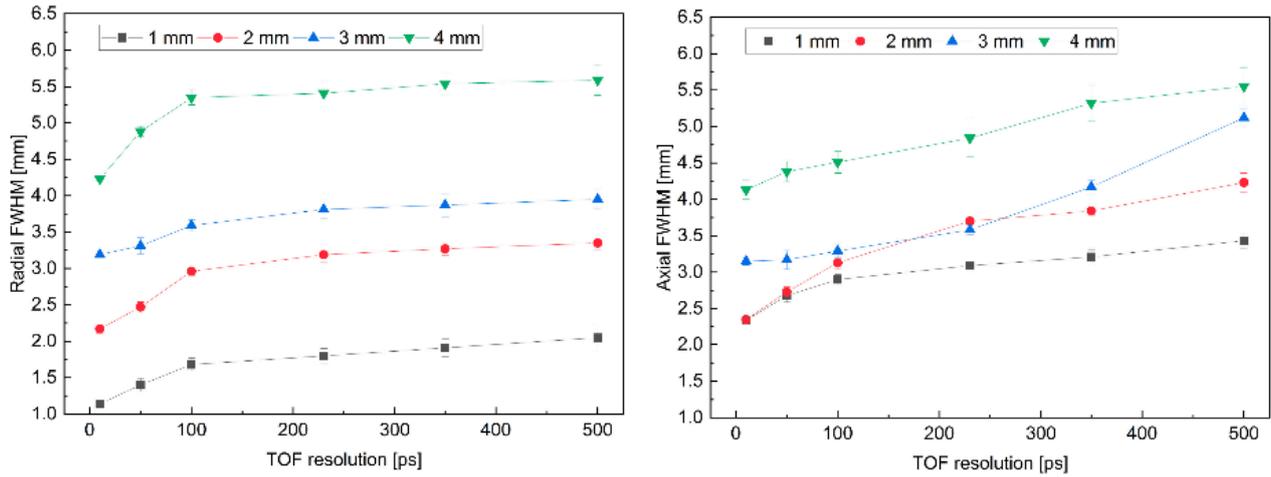

Fig. 4. Plots show the obtained FWHM for various dimensions of plastic scintillator for numerous values of the TOF resolutions. As shown in the plots, there is a visible dependence between achieved FWHM for different plastic scintillator dimensions and corresponding TOF resolution.

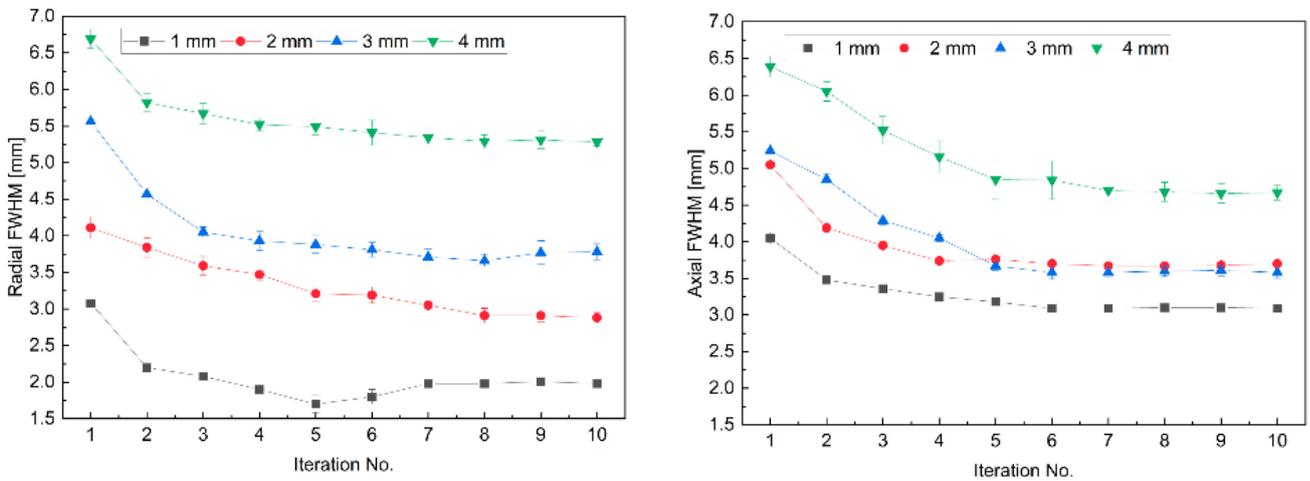

Fig. 5. Achieved FWHM in radial (upper) and axial (lower) as function of number of the iteration for various dimension of plastic scintillator with a point sources at the center of the scanner.

In order to calculate the sensitivity of mouse and rat TB J-PET, a group of point sources (see Table III) was simulated. The obtained values are presented in Fig. 6. In both cases, the central area of the field-of-view is the most sensitive with 2.35% and 2.6% system sensitivity for mouse and rat tomographs, respectively. While moving to the furthest point in the axial direction there is a total decrease by factors of 2.3 and 3.7. The reason for this discrepancy between scanners lies in the difference in their diameters, which directly corresponds to their solid angle coverage over the outermost source.

The spatial resolution of the tomographs has been evaluated as the FWHM of multiple point sources reconstructed with the OSEM algorithm in radial, tangential, and axial directions. Additionally, the effect of the number of the iteration has been estimated. For both investigated systems by increasing the number of iterations, it was possible to reduce resolutions and find their minimal achievable values (see Figs. 7 and 8). In most cases, the plateau region or the local minimum was reached between 6th and 8th iteration. The volumetric resolution refers to the multiplication of spatial resolution values in radial, tangential, and axial directions. For the calculation of the volumetric resolution, the 6th iteration of the centrally located source was chosen (see Table IV). Obtained values were equal to $9.46 \pm 0.54$ mm$^3$ and $14.11 \pm 0.80$ mm$^3$, for mouse and rat TB J-PET, respectively.

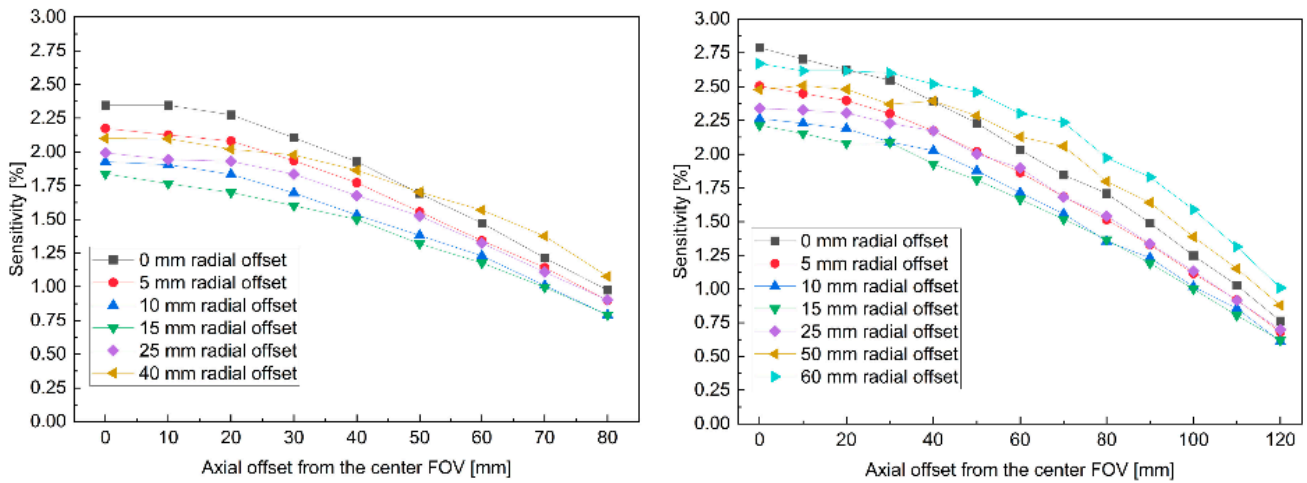

Fig. 6. Absolute sensitivity of mouse TB J-PET (up) and rat TB J-PET (down) as a function of radial and axial offset of sources. Due to the symmetrical configuration of scanners 54 and 65, point sources are able to map the sensitivity of all parts of scanners, respectively.

Alongside offering many advantages utilization of the large AFOV in total body tomographs have a degradable influence on the spatial resolution. It is caused by the contribution of oblique LORs in image reconstruction. In order to solve this problem, an acceptance criterion is being used. Based on the simulation of the centrally located point source of 1-MBq activity, the effect of $45°$ acceptance angle cut on radial, tangential, and axial resolution was estimated (see Table IV). For the calculation of the FWHM-s, the 6th iteration has been used. As can be seen in Table IV, the effect of applying the acceptance angle criterion is negligible. In view of that fact, the acceptance parameter seems to be neutral for both mouse and rat TB J-PET systems.

TABLE IV
SPATIAL RESOLUTION OF SMALL ANIMAL TB-JPET SCANNERS WITH AND WITHOUT ACCEPTANCE CRITERION

| TB J-PET | Mouse | | Rat | |
|---|---|---|---|---|
| Acceptance cut | None | 45° | None | 45° |
| Radial FWHM [mm] | 1.80 ± 0.10 | 1.65 ± 0.02 | 2.05 ± 0.10 | 1.93 ± 0.02 |
| Tangential FWHM [mm] | 1.70 ± 0.02 | 1.65 ± 0.02 | 2.08 ± 0.04 | 1.96 ± 0.02 |
| Axial FWHM [mm] | 3.09 ± 0.03 | 2.95 ± 0.04 | 3.31 ± 0.07 | 3.16 ± 0.02 |
| Volumetric Resolution [mm$^3$] | 9.46 ± 0.54 | 8.03 ± 0.18 | 14.11 ± 0.80 | 11.95 ± 0.19 |

The neutrality of this characteristic in oppose to its impact on human-grade total body scanners can originate from the novel geometrical configuration of plastic scintillator strips. Multiple layers of fine strips in a radial direction within each detection module as shown in Fig. 3 allow for more precise localization of annihilation photon's interactions in the transverse plane, which minimizes the impact of the parallax error.

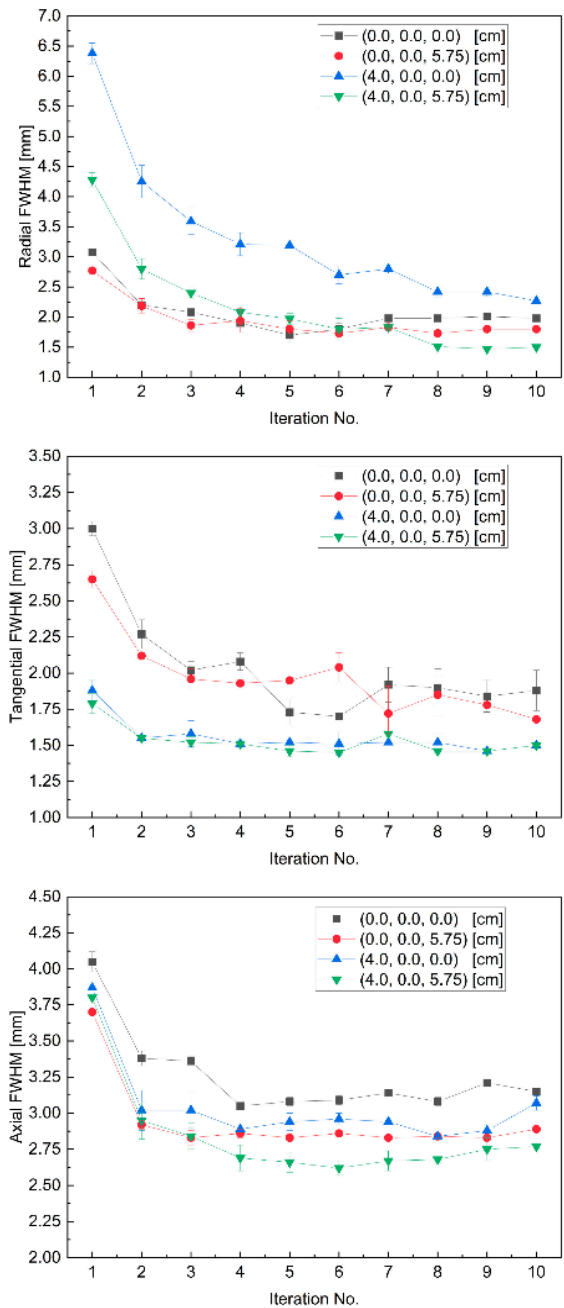

Fig. 7. Plots show the obtained FWHM values for mouse TB J-PET in radial, tangential, and axial axes as the function of the iteration number.

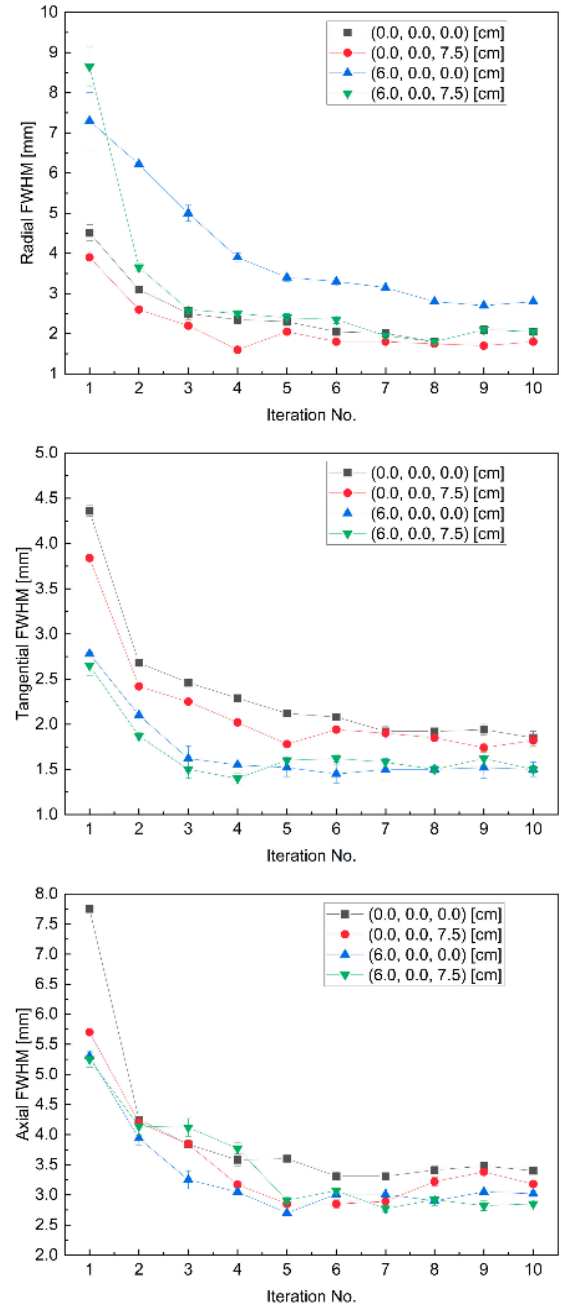

Fig. 8. FWHM values of the reconstructed images of point sources in rat TB J-PET in radial, tangential, and axial axes.

## IV. CONCLUSION

The development of total body PET technology is currently one of the main directions in the medical imaging field. While being mostly pursued in the human-grade tomographs, such technology has great application in small animal studies as well. The main aim of this study was to investigate the feasibility of the small animal total body systems designed with the J-PET technology and based on the simulation performed with GATE software.

Since the scintillator dimension has a direct influence on the spatial resolution, which is the key parameter in rodent studies, its optimal size was investigated. For that four mouse TB J-PETs constructed from scintillator strips of different cross sections were simulated. The presented results show that when it comes to the system sensitivity there is no significant change with varying strip dimensions as long as the total amount of scintillator material remains constant (see Fig. 3). However, as expected, there is an evident impact of the strips dimensions on the spatial resolution. In that case, the best performing cross section is the smallest one, 1 mm × 1 mm.

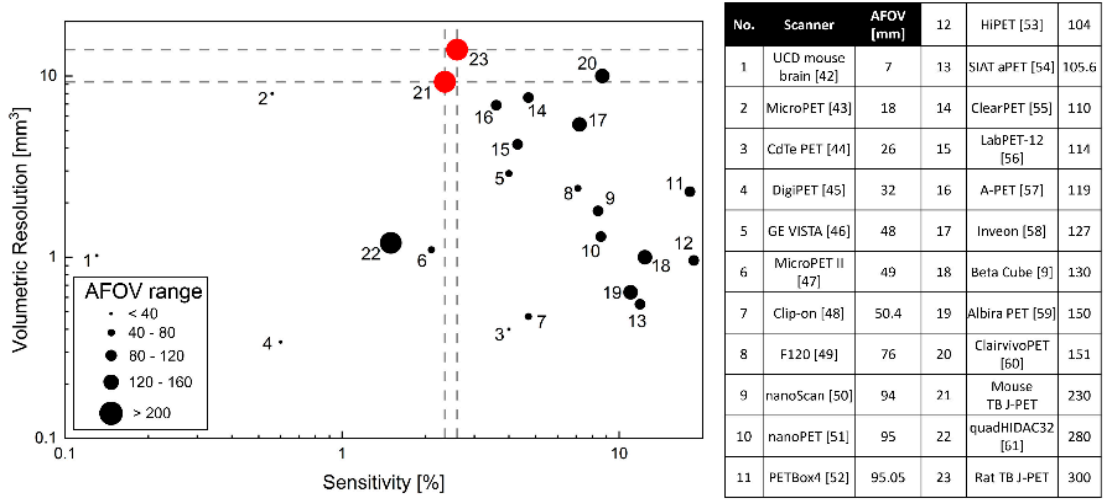

Fig. 9. Comparison of sensitivity and volumetric resolution of preclinical PET systems. TB J-PET solutions are indicated in red (21 for mouse) and (23 for rat). Dashed lines indicate values for preclinical TB J-PET. Numbers indicate results for PET systems as explained in the legend in the right panel of the figure. The size of symbols is proportional to the AFOV of the scanner.

By considering the similarity in the obtained sensitivity, this dimension was chosen for the final design and further study of mouse and rat TB J-PET tomographs.

In order to investigate their performance in detail, mainly the sensitivity and spatial resolution have been used. For the proposed scanners 2.35% and 2.6% of the system, sensitivity is achievable for mouse and rat TB J-PET, respectively. In the same order, they provide 9.46 ± 0.54 mm$^3$ and 14.11 ± 0.80 mm$^3$ of volumetric resolution. Moreover, the impact of the angular acceptance criterion, which is a commonly used parameter to enhance the scanner's resolution, was estimated. The acceptance cut was shown to have a negligible influence on the spatial resolution. This result comes from the unique configuration of the proposed small animal TB J-PETs, mainly the multilayer division of fine scintillator strips in the radial direction within a detection module. Therefore, the designed total body tomographs are almost parallax-free systems.

Finally, the extended AFOV of investigated J-PET scanners not only provides a larger detection area but also the possibility of performing single bed position scans of full, larger size animals. In order to compare the obtained small animal PET tomograph's performance, a comparison plot to some currently available scanners is shown in Fig. 9. This plot is an updated version of the one presented by Lai et al. [63]. Presented results in Fig. 9 come from both iterative and analytic algorithms of image reconstruction. While having the comparable performance, the proposed total body tomographs based on the J-PET technology are able to provide the largest AFOV among all inspected scanners. Alongside cost-effectiveness, this solution is a feasible choice for preclinical studies on small animals.